\documentclass[sigconf,natbib=true,anonymous=false,authorversion=true,review=false,pbalance=true]{acmart}

\usepackage{amsmath}%
\usepackage{rotating}%
\usepackage{color,soul}
\usepackage{tabu}%
\usepackage{multirow}%
\usepackage{acronym}
\usepackage{makecell}
\usepackage{enumitem}

\setlist{parsep=.1em,itemsep=.1em,topsep=.1em,leftmargin=1em}

\newcommand{\ld}[1]{}

\newcommand{\car}{TREC CAR Y3}
\newcommand{\tqa}{TQA}
\newcommand{\dl}{TREC DL}
\newcommand{\dlfirst}{TREC DL 2019}
\newcommand{\dlsecond}{TREC DL 2020}
\newcommand{\genQ}{GenQ}

\newcommand{\questionanswering}{quest\-ion ans\-wer\-ing}
\newcommand{\irsystems}{ret\-rie\-val gen\-era\-tion sys\-tems}
\newcommand{\irsystem}{ret\-rie\-val gen\-era\-tion sys\-tem}

\newcommand{\examcover}{EXAM Cover}
\newcommand{\examqrels}{EXAM Qrels}
\newcommand{\treceval}{\texttt{trec\_eval}}
\newcommand{\squad}{SQuAD2}

\newcounter{UseArxiv}
\begin{document}
\ifnum\value{UseArxiv}=1
\title{An Exam-based Evaluation Approach \\ Beyond Traditional Relevance Judgments}
\else
\title{Pencils Down! Evaluating Information Content with FLAN-T5-EXAM}
\fi
%\title{EXAM++: Exam Answerability Metric with LLMs}
% \title{EXAM++: Evaluating with Exam Questions instead of Relevance Judgments}

\author{Naghmeh Farzi}
\email{Naghmeh.Farzi@unh.edu}
% \orcid{\ld{needed!}}
\affiliation{%
  \institution{University of New Hampshire}
  \country{USA}
}
% \author{Jaekeol Choi}
% \email{todo}
% \orcid{\ld{needed!}}
% \affiliation{%
%   \institution{University of New Hampshire}
\author{Laura Dietz}
\email{dietz@cs.unh.edu}
% \orcid{\ld{needed!}}
\affiliation{%
  \institution{University of New Hampshire}
  \country{USA}
}

\ifnum\value{UseArxiv}=1

\begin{abstract}
    Current IR evaluation is based on relevance judgments, 
    created either manually or automatically, with decisions outsourced to Large Language Models (LLMs). We offer an alternative paradigm, that never relies on relevance judgments in any form.  Instead, a text is defined as relevant if it contains information that enables the answering of key questions. 
    We use this idea to design the EXAM Answerability Metric to evaluate information retrieval/generation systems for their ability to provide topically relevant information.
        
    We envision the role of a human judge to edit and define an exam question bank that will test for the presence of relevant information in text. We support this step by generating an initial set of exam questions.
    In the next phase, an LLM-based \questionanswering{} system will automatically grade system responses by tracking which exam questions are answerable with which system responses. We propose two evaluation measures, the recall-oriented \examcover{} metric, and the precision-oriented \examqrels{} metric, the latter which can be implemented with \treceval{}. This paradigm not only allows for the expansion of the exam question set post-hoc but also facilitates the ongoing evaluation of future information systems, whether they focus on retrieval, generation, or both.
    % The paradigm allows to expand the exam question set post-hoc and to evaluate future information systems, based on retrieval and/or generation.
    \footnote{Data and code available at \url{https://github.com/laura-dietz/flan-t5-exam-appendix}}
    
\end{abstract}

\else

\begin{abstract}
Large language models and retrieval-augmented generation (RAG) are challenging traditional information retrieval evaluation methods. System responses should cover relevant information content while being concise and fluent. However, current evaluation metrics fall short in accurately assessing the information content of systems' responses---without resorting to expensive human judgments.

In contrast, the EXAM Answerability Metric puts information retrieval systems to the proverbial test. This metric leverages a bank of query-related exam questions 
to quantify relevant information content that is covered in the systems' responses.
The process involves (1) decomposing the query into detailed questions, and (2) checking each for answerability using passages in the system response.
Using two TREC benchmarks, we demonstrate that our LLM-based EXAM approach works successfully in both of these phases. Nevertheless, manual verification can ensure that the questions test important aspects of the information need, and to verify the answer checking stage.\footnote{Data and code available at \url{https://anonymous.4open.science/r/flan-t5-exam}}

\end{abstract}

\fi

% \keywords{Information Retrieval Evaluation}

\maketitle

%  ACRONYM Definitions
% use with \ac{LLM} or \acp{LLM}
\begin{acronym}
    \acro{LLM}[LLM]{Large Language Model}
\end{acronym}

\section{Introduction}
\label{sec:introduction}

 \acp{LLM} are capable of generating and/or retrieving responses for search queries, resulting in a plethora of systems that combine traditional retrieval with neural ranking and natural language generation.
 Ideally the systems' responses cover relevant information content while being concise and complete. However, there is a  need for convincing evaluation metrics to assess the accuracy and completeness of the information content in responses. This should be accomplished in a repeatable and reusable manner and without resorting to expensive human judgments.

In the remainder of the paper we will focus on the evaluation task of the following process:

\noindent 
\paragraph{\textbf{Task Statement.}}
An information retrieval/generation system is given a search \emph{query} to produce a relevant \emph{system response}. The response can take the form of a passage ranking, a set of extractive summaries, or a single (generated) text. 
\\
Given system responses across multiple queries from multiple systems, the task is to assign each system an \emph{evaluation score} that represents the quality of the information content provided in the system response.

\bigskip

A popular evaluation approach is to ask \acp{LLM} whether a passage is relevant for a query. Empirically this has been shown to work well \cite[\textsl{--inter alia}]{faggioli2023perspectives,macavaney2023one,thomas2023large}. 
However, skepticism remains about whether \acp{LLM} can be trusted to completely replace humans in the judgment process \cite{faggioli2023perspectives}.

A solution may be to integrate a human oversight into the evaluation process. While using human judges to verify automatic relevance labels is a common suggestion, it can be a tedious task without additional context.

. The common suggestion is to employ judges to verify automatic relevance labels. However, without further context, this remains a tedious task.  While LLMs can be asked to produce a rationale to support their labeling decision,  studies indicate that human judges might overly trust rationales generated by LLMs \citep{fok2023search}.

Acknowledging that judging passage-level relevance is a daunting task---being simultaneously cognitively challenging and boring---we propose a more ``human-friendly'' way to leverage human labor for defining what constitutes relevant information content. In this paper we focus on the empirical evaluation of an alternative automatic approach, leaving human-subject studies for future work.

\begin{figure*}
    \includegraphics[width=0.72\textwidth, height=0.42\textwidth]{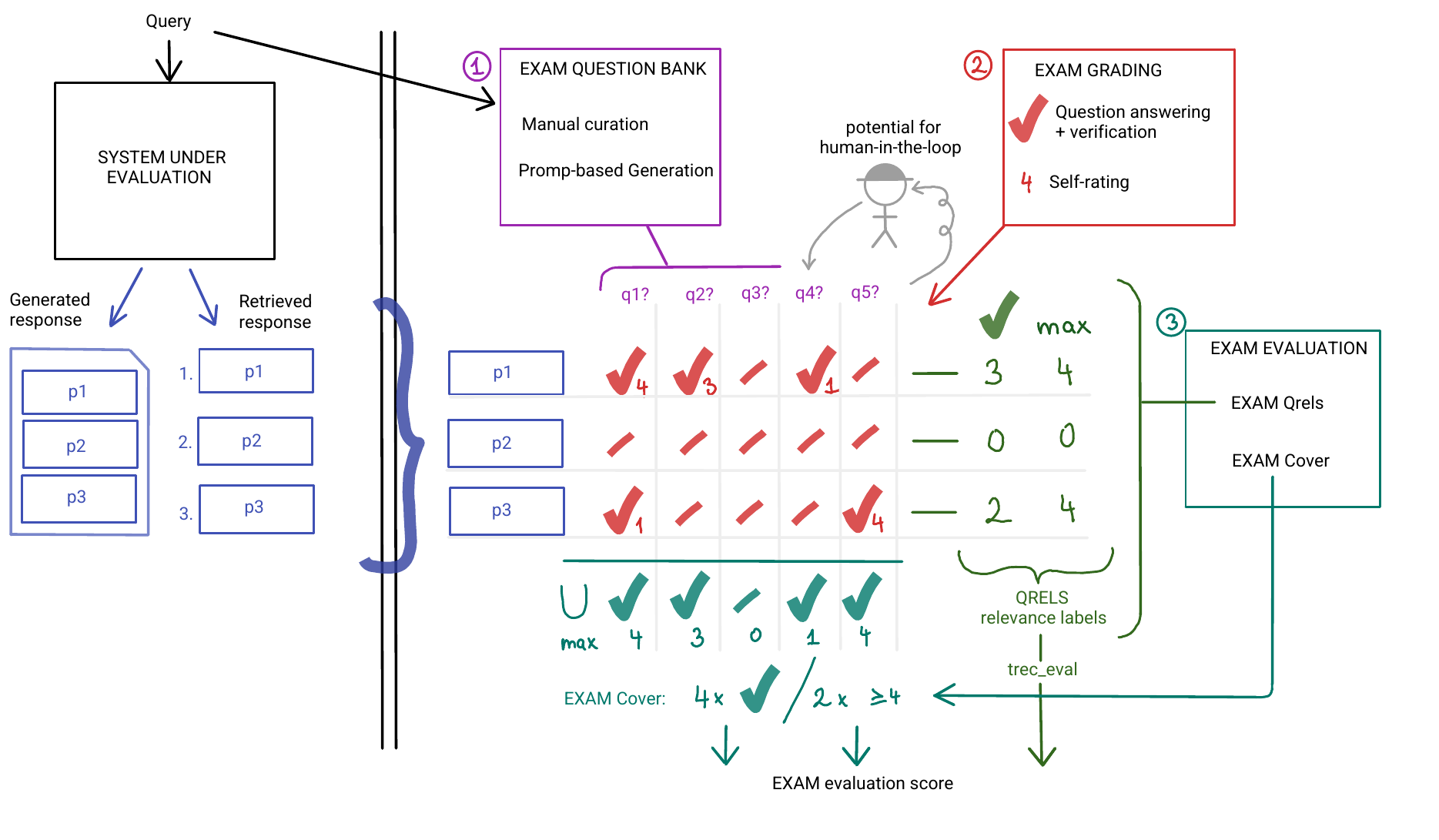}
    \caption{EXAM approach. Left: The only information available to the system that is being evaluated are queries and,  optionally, a text corpus. The system response can be a ranking of passages, or a generated response that will be segmented into passages (\textcolor[RGB]{0,20,150}{blue}). Passages from all systems will be pooled for assessment, and additional passages can be added at a later if needed. Right: The EXAM evaluation system uses three phases detailed in Section \ref{sec:approach}. For each query, an exam question bank is developed, which can be modified later in an iterative fashion (\textcolor[RGB]{100,0,70}{purple}). All passages from the system response (e.g., p1, p2, p3) are graded based on which questions (q1?, q2?, ..., q5?) can be correctly answered with the passage text (\textcolor{red}{red}). We support two modes: one where answers are verified against an answer key (depicted as check marks), or by having an LLM self-rate the answerability on a scale from 0 to 5. The EXAM evaluation scores are derived from these grades (\textcolor[RGB]{0,90,50}{green}). The \examcover{} score is based on how many questions are covered, as binary verification or via a minimum self-rating level. For \examqrels{} a relevance file for \treceval{} is derived, which is based on the coverage or best self-rating obtained by this passage in isolation. We provide a worked example in Section \ref{sec:worked-example}\label{fig:approach}}
\end{figure*}

 Our goal is to develop an evaluation approach that,
 \begin{enumerate}
     \item is amenable to integrating a human into-the-loop in a less daunting way, \label{en:human-in-loop}
     \item never requires manual passage-level relevance judgments, \label{en:no-assessments}
     \item benefits from latest advances in \aclp{LLM}\label{en:llm-benefits}
     \item yields reusable test collections that can  evaluate (future) systems, especially those that employ language generation \label{en:reusable}
     \item allows to expand the test collection post-hoc to reveal differences between systems \label{en:expandable}
 \end{enumerate}
Of course, such an evaluation approach has to agree with traditional evaluation paradigms, as quantified by rank correlation of system leaderboards.% inter-annotator agreement and .

\bigskip

%\ld{make more clear what difference between EXAM and us is}

%\ld{make clear relationship between query and questions}

We propose that, for every query, information \irsystems{} take an exam for which they did not study. Using a query-specific question bank of exam questions, the EXAM Answerability Metric determines whether important information content is provided in the system response to that query. 

As part of topic development, exam questions need to be designed for each piece of desired information content. Whenever it is possible to answer this question with the system's response, we consider this piece of information content covered.

The idea was first introduced by \citet{sander2021exam}, who envision that (1) human judges would manually create the exam question bank, while (2) an automated \questionanswering{} system performs the daunting task of grading each system response. The system's EXAM evaluation score (3) is the higher, the more questions can be correctly answered based on the system's response.

\paragraph{Contributions.} We significantly expand on the EXAM idea with our proposed approach FLAN-T5-EXAM, depicted in Figure \ref{fig:approach}.

\begin{itemize}
    \item FLAN-T5-EXAM supports development of exam question banks with prompt-based generation,
    \item modernizes the \questionanswering{} system with the recently released FLAN-T5 family \cite{longpre2023flan},
    \item proposes an EXAM-variant that is inter-operable with commonly used evaluation tools (e.g. \treceval{})
    \item expands the empirical evaluation to additional datasets from \dl{} 2019 and 2020 and baselines from automated relevance prompting \cite{sun2023chatgpt,faggioli2023perspectives,thomas2023large}.
\end{itemize}   

A strength of this approach is that, in contrast to work LLM-based relevance grading, we can readily integrate humans into the evaluation (Goal \ref{en:human-in-loop}). Based on the long history of classroom education and exams, we argue it is more natural for human judges to control the design of exam questions than to directly provide relevance judgments. These could be developed manually from scratch, on the basis of system responses, or in dialog with a chat-LLM. In this work we focus on the two extreme cases: fully manually created exam questions and fully automatically generated exam questions---leaving human subject studies to future work.

% \ld{explicit about the compromise}
% While out-of-scope for this work, we imagine that humans should be incorporated in the creation of exam question banks, by inspecting and modifying the initial question bank. This could be performed manually, on the basis of  system responses, or in dialog with a chat-\ac{LLM}. Furthermore, exam questions can be prioritized according to a multi-level relevance scale such as ```must have answered'', ``should have answered'', or ``would be nice to have answered''.

By virtue of automating the grading of system responses, human judges are never required to perform passage-level relevance assessments (Goal \ref{en:no-assessments}). The evaluation approach yields reusable test collections (Goal \ref{en:reusable}) that can be expanded by modifying the question bank (Goal \ref{en:expandable}) at any point in the evaluation process, as the remaining pipeline is fully automated. The impact of a question bank modifications can be directly observed by listing passages whose relevance grade would change.

At the same time, humans are fully in control of defining which information content is relevant via the exam question bank. 

\section{Related Work}
\label{relatedwork}

We focus on an approach that does not require passage-level relevance judgments or source texts. Our work is unique in this regard, but aspects relate to many active branches of research, which we detail below.

% in any form, and which does not intend to mimic this relevance-judgment process.  Instead the evaluation is based on correctly answering questions with system responses, using an LLM-based approach. 

% \ld{Also, no gold summary is ever needed.}

\subsection{LLM-based Relevance Assessment}

While our approach does not attempt to imitate the relevance-judgment process, several recent methods studied this approach.

 \citet{sun2023chatgpt}  %\url{https://arxiv.org/pdf/2304.09542.pdf}
 rerank passages using a simple LLM prompt ``does the passage answer the query?'' We use their prompting technique as a baseline ``\texttt{Sun}'' for comparison.

\citet{faggioli2023perspectives} conduct an early experiment by asking an LLM to judge the relevance of a passage. They design a simple prompt ``\texttt{Fag}'' and a more elaborate multi-relevance few-shot prompt developed for \dl{}. 

\citet{thomas2023large} compare the ability of LLMs to perform document-level relevance judgments in comparison to different groups of human annotators. They find that especially the label quality of crowd-workers is inferior to fully automatic LLM-based relevance labels. \citeauthor{thomas2023large} are using a very detailed prompt (Figure 1 in \cite{thomas2023large}), clarifying the role and query description and asking the LLM to comment on the query intent and trustworthiness. As the full prompt well exceeds the 512 token boundary of FLAN-T5-large models, we use an abridged prompt ``\texttt{Thom}''.\footnote{All baseline prompts are provided in our online appendix.}
  
In 1SLs, \citet{macavaney2023one} focus on evaluating passages with a DuoPrompt, that instructs an LLM to indicate which of two passages is more relevant for a query. In contrast, our approach does not fall in the paradigm of pair-wise ranking approaches.
\bigskip

Several voices raised critiques about using LLM's for relevance labels in general. \citet{faggioli2023perspectives} elaborates a wide-range of theoretical concerns, centered on questions of trustworthiness and reliability of LLMs now and in the future. \citet{wang2023large} empirically demonstrate that LLMs exhibit unfair positional bias towards candidates displayed for evaluation. \citet{fok2023search} studies general issues of human over-reliance and under-reliance on LLMs. They elaborate why rationales produced by LLMs for human verification do generally not lead to improvements.

\subsection{Nugget-based Evaluation}

There is a long history in the IR community to evaluate the relevance of documents by breaking down the information need into a set of ``nuggets'' (also called query intents, facts, or SCUs) that can each be evaluated independently \citep{lin2006will}. The common definition of a nugget is ``the smallest portion of text that constitutes relevant information in and of itself'' \cite{pavlu2012ir}. While text is defined as relevant as soon as it matches any nugget, the nugget document score is based on the sum of shingle match scores in any document---to be considered as relevant when surpassing a threshold.

% \url{http://idl.iscram.org/files/mccreadie/2023/2529_McCreadie+Buntain2023.pdf}
With the advent of LLMs, nugget-style evaluation is being revamped, most recently in the TREC Crisis Facts track \cite{mccreadie2023crisisfacts}: Judges are asked identify atomic ``facts'' (similar to nuggets). System responses are analysed for mentions of these facts, either via a boolean OR or with an embedding-based method.

Our proposal is related in that our questions are intended to capture nuggets. A hurdle in using nugget-based evaluation method is to reliably match nuggets to text content. In contrast, with FLAN-T5-EXAM, we are leaning on decades-long research of factoid \questionanswering{}---which is now considered a solved task.

\subsection{Evaluation with Test Questions}\label{sec:rw-test-questions}

% \citet{sander2021exam} proposed the EXAM Answerability Metric for the evaluating information retrieval systems that respond to under-specified information needs \ld{demanding wide coverage}. We draw inspiration from this work, and delineate our proposed approach in Section \ref{sec:related-work-difference-sander} how our work differs. %We include this method in our experimental evaluation in Section \ref{sec:evaluation}.

The idea of basing an evaluation on a bank of test questions has been widely discussed in literature on summarization \cite{clarke2010discourse}, recently with automated \questionanswering{} methods. 
\citet{eyal-etal-2019-question-summary-first} suggest a system evaluation score that is based on the number of questions that a \questionanswering{} system can correctly answer using the system response---a principle that both the original EXAM method and our approach follows.

In contrast to our approach,
%\citeauthor{eyal-etal-2019-question-summary-first} and \citet{scialom-2019-answers-bert} automatically generate questions from a ground truth source text. 
most approaches use a Cloze-style idea to generate questions from a given gold summary or source text.  Questions can be in the form of multiple-choice questions \cite{Huang2020qasummary}, free text questions with exact-match answer verification \cite{deutsch2020questionanswering}, or be derived from extracted entities and relations \cite{wang-etal-2020-factual-qa,eyal-etal-2019-question-summary-first}. 
% %
% \ld{drop?  To improve the quality of generated questions \citet{zhou2018neural} and \citet{murakhovska-etal-2022-mixqg} suggest to train LLMs on how to generate relevant questions leveraging several available datasets, such as SQuaD, NewsQA, BoolQ, HotpotQA.
% %
A full survey is provided by \citet{mulla2023automatic}.

As it pertains to information retrieval evaluation, the problem with generating questions from a given text or gold summary, is that (1) such a gold summary is usually not available and (2) it is unclear which of these questions relate to \textsl{relevant information} in the gold summary (or source text).

%
% \ld{drop?
% While \citet{scialom2021questeval} include a question weight, they suggest to determine this weight based on the gold summary. } ]
The original EXAM method avoids this problem altogether by asking a human to design questions that address the search query.
In contrast, we propose to automatically generate questions directly from the query, building on the world-knowledge of ChatGPT---with the intention of employing manual labor to verify or weight the question set.

\subsection{LLMs, Passages, and Questions}

Many approaches integrate passages, questions, and LLMs in some form. This includes improving \questionanswering{} via retrieval-augmentation \citep{sachan2022improving,li2023llatrieval}.
Improve factuality in fact verification, by breaking each claim down into several questions \citep{zhang2023towards}.
Exploiting the self-verification ability of LLMs to improve the reasoning \citep{weng2023large}.
Evaluating the quality of LLMs with multiple tests  \citep{liang2022holistic,chang2023survey}.

\citet{arabzadeh2024adapting} develop an approach to evaluate retrieved passages via questions in a very different way: For search queries that resemble questions (as in \dl{}), LLMs are asked to generate a correct answer and an incorrect ``liar'' answer. These responses as used as positive/negative gold answers for retrieved passages: If the embedding-based similarity of the passage is closer to the correct answer than the incorrect answer, the passage is deemed relevant. 
%
% This falls in a class of approaches that \ld{derive representations of correct and  incorrect poles, then classify by the one that is more similar.}
% identify representations for correct-vs-incorrect or relevant-vs-nonrelevant information \cite{stoehr2023unsupervised}. 

% Naghmeh's summary
% In [Negar's paper] the authors conducted research where they employed the MS MARCO V1 passage retrieval dev set, TREC Deep Learning (DL) 2019, and 2020 to evaluate the generated answers of the large language models (LLM). They utilized GPT models to generate answers and intentionally prompted models to produce incorrect responses as well. They used cosine similarity to assess the similarity between generated and relevance-judged retrieved passages having a high correlation. The study also evaluated the similarity between the generated answers and top-retrieved passages from various retrieval models.

% \begin{verbatim}
%     - We focus on a setting where no passage-level relevance judgment is ever needed. Also, no gold summary is ever needed.

%     How to match correct responses?  

%    Quotes for "LLMs are unreliable" 
%    * Bowman :Eight Things to Know about Large Language Models" \url{https://arxiv.org/pdf/2304.00612.pdf}
%    * E. Ferara: Should chatgpt be biased? challenges and risks of bias in large language models
% \end{verbatim}

% \ld{ --- include this! ----- }

\subsection{Difference to Original EXAM \citep{sander2021exam}}

% The original EXAM method relies solely on humans to design the exam, with the intent that only a human could identify the core questions that would need to be addressed in a relevant answer. This is in contrast to approaches that generate questions from a gold summary (detailed in Section \ref{sec:rw-test-questions}), which might lead to questions derived from non-relevant aspects mentioned in relevant text.
% In contrast, our approach directly generated an initial set of query-relevant questions via a generative LLM (such as chatGPT \cite{chatgpt}). 

Our contribution differs from the original EXAM method in several important ways:

% \ld{change layout to flowable text}
\begin{description}
    \item[1. Obtaining an Exam Question Bank:] To obtain  exam  questions,
    \begin{itemize}
        \item The original EXAM method is based on manually created multiple-choice exam questions.
        \item We propose to semi-automatically generate free-text questions for each query, as described in Section \ref{sec:generating-question-banks}.
    \end{itemize} 
    \item[2. Grading System Responses:]  To grade each passage via the answerability of exam question,
    \begin{itemize}
        \item The original EXAM method uses a pre-neural multiple-choice \questionanswering{} system with answer verification. 
        \item First, we propose to modernize the \questionanswering{} system with an LLM-based approach (Section \ref{sec:exam-grading-answer-verification}).% While any LLM can be used, we experiment with the recent FLAN-T5 family,  with and without SQUAD2 fine-tuning.
        \item Second, we explore the ability of LLMs to self-rate the answerability of a question with given context, without directly verifying the correctness of answer (Section \ref{sec:exam-grading-self-rating}).
    \end{itemize}
    \item[3. \examcover{} Evaluation:] To evaluate each system, 
    \begin{itemize}
        \item With \examcover{}, we follow the original EXAM method by evaluating systems according to the number of answerable exam questions (Section \ref{sec:evaluation-exam-cover}).
        
        % \begin{equation}
        %     \text{\examcover{}}(P)=\frac{1}{ \|Q\|} \| \bigcup_{p\in P} \{q | \text{correct}(q,p), \forall q\in Q\|
        % \end{equation}
        % In the case where some exam questions are overly difficult, the \examcover{} score can be normalized by the questions that are answerable by any system or a gold system response.
        \item To improve adoption, we add a variant ``\examqrels{}'' that implements a related idea so that it is inter-operable with the popular evaluation tool \treceval{} (Section \ref{sec:evaluation-exam-qrels}).
    \end{itemize}
\end{description}

\section{Approach}
\label{sec:approach}

The \examcover{} metric quantifies the information content of retrieved, generated, or retrieval-augmented generated information.
Using a question bank of exam questions, the systems are graded based on the set of exam questions that can be correctly answered based on the system's response: The more questions can be answered, the higher the EXAM score of the system.

Our EXAM evaluation system assumes the following inputs:
\begin{enumerate}
    \item A set of queries, optionally with query subtopics.
    \item A set of system responses, which can come in the form of a passage ranking or as generated content.
\end{enumerate}

Note that the exam questions are secret, and not available to the \irsystem{}.

The EXAM evaluation approach is structured into the following three phases that we detail in the remainder of the section, and depict in Figure \ref{fig:approach}. 
\begin{description}
    \item[1. Obtaining an Exam Question Bank:] A process of creating an exam of questions, with or without gold standard answer keys. 

    \item[2. Grading System Responses:]  All passages in system responses are graded using an automated \questionanswering{} system. The set of answerable questions is tracked for each passage.
  
    \item[3. EXAM Evaluation:]  The system's evaluation score that is the higher, the more exam questions can be answered with any of the first $k$ passages of the system's response.
\end{description}

\begin{table*}[ht]
    \caption{ Question generation prompts (\genQ{}). \label{tab:prompt-table}}

\begin{tabular}{p{0.45\linewidth} p{0.45\linewidth} }
    \toprule
    \textbf {Question Generation: TREC DL Prompt} & \textbf{Question Generation: TREC CAR Y3 Prompt} \\
    % \midrule
    
     Break the query '\{query\_title\}' into concise questions that must be answered. Generate 10 concise insightful questions that reveal whether information relevant for '\{query\_title\}' was provided, showcasing a deep understanding of the subject matter. Avoid basic or introductory-level inquiries. Keep the questions short and in a Python list format. & Explore the connection between '\{query\_title\}' with a specific focus on the subtopic '\{query\_subtopic\}'. Generate insightful questions that delve into advanced aspects of '\{query\_subtopic\}', showcasing a deep understanding of the subject matter. Avoid basic or introductory-level inquiries. Give the question set in a Python list format.\\
    \bottomrule
  \end{tabular}
\end{table*}

\begin{table}[h]
    \caption{ Grading prompts. \label{tab:qa-prompt-table}}

\begin{tabular}{p{0.95\linewidth} }
    \toprule
    \textbf {Grading: Question Answering Prompt}  \tabularnewline
    % \midrule
    % \makecell{
    \parbox[t]{0.95\linewidth}{
    provide a complete and concise answer to the question based on the context.
    Question: \{question\}
    Context: \{context\} 
    }\tabularnewline
    \midrule
    % \medskip
    \textbf {Grading: Self-rating Prompt}  \tabularnewline
    % \midrule
    % \makecell{
    \parbox[t]{0.95\linewidth}{
Can the question be answered based on the available 
context? choose one: \\
- 5: The answer is highly relevant, complete, and 
accurate.\\
- 4: The answer is mostly relevant and complete but may
have minor gaps or inaccuracies.\\
- 3: The answer is partially relevant and complete, with 
noticeable gaps or inaccuracies.\\
- 2: The answer has limited relevance and completeness, 
with significant gaps or inaccuracies.\\
- 1: The answer is minimally relevant or complete, with 
substantial shortcomings.\\
- 0: The answer is not relevant or complete at all. \\
Question: \{question\}
Context: \{context\} 
}\tabularnewline
    \bottomrule
  \end{tabular}
\end{table}

\subsection{EXAM Question Banks}

\subsubsection{Generating EXAM Question Banks (\genQ{})} \label{sec:generating-question-banks}
We employ a generative LLM to automatically generate free-text- questions for each query.

In our experiment, we use ChatGPT to obtain question banks. The prompt is designed to elicit concise, insightful questions based on the query, tailored to specific goals of the IR task and domain. In application to \car{}, we ask for questions that explore the connection between the query with a specific focus on each subtopic. For \dl{} we ask to break the query into concise questions. The complete prompts used in the experimental evaluation are listed in Table \ref{tab:prompt-table}.

We obtain free-text questions without answer key from these prompts, such as the following for \car{} query  \texttt{tqa2:L\_0384}:

\begin{quote}
\textbf{Query title:} The Integumentary System \\ \textbf{Query subtopic:} Structure of the Skin
\begin{itemize}
    \item[q1?] How does the epidermis, dermis, and hypodermis work together to provide protection, sensation, and regulation for the body? 
    \item[q2?] Can the integumentary system be compromised by diseases and conditions, and if so, how does this impact the health of the skin?
    \item[q3?] How does the skin act as a barrier against pathogens and other foreign substances?
\end{itemize}
\end{quote}

\subsubsection{Manual Verification and Refinement of EXAM Question Banks} \label{sec:manual-verification}

Our method facilitates the inclusion of a human-in-the-loop when constructing the exam question bank. For example, the human judge could verify that the question bank is covering essential information content for the query. Questions can be changed, added, or deleted at any time: during topic development and during a post-evaluation analysis. As better systems are being developed, more difficult questions can be added to reveal quality differences between systems.

While our evaluation system is designed to support the human-in-the-loop, we leave the user study to future work.

\subsubsection{Manual Question Banks}\label{sec:manual-question-banks}
Alternatively, question banks can be created from scratch, as suggested for the original EXAM method. If  questions are associated with a gold answer key, it can be used for answer verification. However, we acknowledge that the requirement of an answer key renders the topic development more difficult.
\texttt{TQA} collection \cite{Kembhavi2017TQA} provides an exam question bank of multiple-choice tests for topics in \car{}.

\subsection{EXAM Grading} \label{sec:exam-grading}

In this phase, we conduct a secret exam using the system responses as background information. We track which of the questions are answerable with which passage.

The original EXAM method uses a pre-neural \questionanswering{} system from AI2, which was designed to answer multiple-choice questions with given context. The answer is verified as correct when the \questionanswering{} system responds with the correct choice. The downside is that this \questionanswering{} system was difficult to set up,\footnote{We were unable to install the \questionanswering{} system used in the original EXAM method.} and was specifically designed for questions in the style of the \texttt{ARC} and \texttt{TQA} datasets. A goal of our work is to modernise this \questionanswering{} system and allow for broader applicability.

\subsubsection{Pre-processing System Responses}

To initialize the grading phase, we pre-process system responses.  Longer responses are segmented into paragraph-sized passages, each approximately 400 tokens in length. In case of submitted passage rankings, each passage will graded individually. We convert each  passage into plain text and assign a unique \texttt{passage\_id} for later use.

\subsubsection{LLM-based \questionanswering{} with Answer Checking} \label{sec:exam-grading-answer-verification}

To take advantage of the latest development in LLMs, we propose to use an LLM-based \questionanswering{} system. While any LLM could be used, in this work we experiment with the FLAN-T5 family that became recently available. In particular we focus on the FLAN-T5-large model which can complete annotations within a reasonable time on an NVIDIA A40 GPU.

\paragraph{Question prompt.}
We perform some initial prompt engineering\footnote{A version of FLAN-T5-large fine-tuned on the \squad{} was also considered.} to ensure that the LLM is answering the question with the provided context, and not from its own memory.
For the FLAN-T5-\questionanswering{} system we use a prompt to instruct the LLM to ``provide a complete and concise answer to the question based on the context''. The full prompt is given in Table \ref{tab:qa-prompt-table}.

Using separate requests to avoid position bias, we obtain answers each exam question on all passages.
Whenever the prompt exceeds the 512 token boundary, we truncate the given context but ensure that the question remains intact.

\paragraph{Answer verification.}
When a known correct answer is available for the question bank, we verify the predicted answer with a heuristic matching function:  Both the correct and predicted answers are normalized by lower-casing, stopword removal, and stemming. A match is considered when the edit distance between these strings is less than 20\% of the length of the longer string. This heuristic has been manually verified to yield accurate matches, effectively reducing false negatives compared to strict exact matching.

\subsubsection{EXAM Grading by Self-rating} \label{sec:exam-grading-self-rating}

Since our focus is on whether a question is answerable within a given context, rather than the answer itself, we explore a way to avoid difficulties in producing gold answers and the verification of predicted answers altogether.  We lean on the ability of modern LLMs to match language patterns and observed self-verification behavior \citep{weng2023large} and ask the LLM self-rate the answerability of a question with the context. We use the prompt given in Table \ref{tab:prompt-table}.

% \begin{verbatim}
% Can the question be answered based on the available 
% context? choose one:
% - 5: The answer is highly relevant, complete, and 
% accurate.
% - 4: The answer is mostly relevant and complete but may
% have minor gaps or inaccuracies.
% - 3: The answer is partially relevant and complete, with 
% noticeable gaps or inaccuracies.
% - 2: The answer has limited relevance and completeness, 
% with significant gaps or inaccuracies.
% - 1: The answer is minimally relevant or complete, with 
% substantial shortcomings.
% - 0: The answer is not relevant or complete at all.
% Question: $question
% Context: $context
% \end{verbatim}

We find that in the vast majority of cases, FLAN-T5 indeed responds with a numerical code between 0 and 5. In the remaining cases, we assign a rating of 1 by default or 0 when expressions of unanswerability are encountered.\footnote{``unanswerable''
,``no''
,``no answe''
,``not enough information''
,``unknown''
,``it is not possible to tell''
,``it does not say''
, or ``no relevant information''.}

\subsection{EXAM Evaluation}

\subsubsection{\examcover{} Evaluation} \label{sec:evaluation-exam-cover}

 We follow the suggestion of \citeauthor{sander2021exam} and \citeauthor{eyal-etal-2019-question-summary-first} to evaluate systems by the number of questions $Q$ that can be answered with any of its passages $P$. 
        
        \begin{equation}
            \text{\examcover{}}(P)=\frac{1}{ \vert Q \vert} \left\vert \bigcup_{p\in P} \left\{q \vert \text{correct}(q,p), \forall q\in Q \right\} \right\vert
        \end{equation}

Depending on the grading method chosen in Section \ref{sec:exam-grading}, for a passage $p$, the question $q$ is considered "correct" when
\begin{itemize}
    \item the answer was verified as correct
    % \item the answer did not entail expressions of unanswerability
    \item the self-rating is above a threshold (e.g. $\geq 1$ or $\geq 4$)
\end{itemize}

The goal of the exam is to assess the comprehensiveness of the information presented in the entire ranking. Hence, \examcover{} monitors the set of questions  that can be accurately answered using any passage in the ranking. \examcover{} constitutes a recall-oriented evaluation metric that does not reward system for providing redundant content. Instead, it encourages to cover the breadth of relevant information content with the passage set $P$.

%Moreover, it is unrealistic to expect a single passage to address all the exam questions.

% \subsubsection{How NOT to use EXAM}

% In evaluating rankings, it might seem intuitive to assess each passage based on its ability to correctly answer questions and then average these scores, similar to NDCG. However, this approach is not advisable.

\subsubsection{\examqrels{} Evaluation: EXAM for trec\_eval}\label{sec:evaluation-exam-qrels}

Recognizing the benefits of integrating novel evaluation metrics into established evaluation tool-chains, we propose a variant of EXAM that can be implemented with \texttt{trec\_eval}.
We can convert EXAM grades to a relevance ``qrels'' file.  We mark any passage as relevant whenever it allows answering an exam question that was generated for a query (or query-subtopic). This yields a binary relevance label for each passage $p$:

\begin{equation}
    \text{relevance-label}(p)=\exists q\in Q : \text{correct}(q,p)
\end{equation}

The ranking can be evaluated with \treceval{} using this \examqrels{} file using any traditional evaluation metric. In the empirical evaluation we use \texttt{Precision@k}, which reflects the proportion of passages in the top $k$ with relevant information content.

This approach avoids the problem of unjudged passages (also called ``holes'' \citep{macavaney2023one}) in test collections. The EXAM grading pipeline can be applied to update the qrels file whenever new passages appear in a system response.

\paragraph{Self-rated Answerability.}

We can extend the \examqrels{} idea to multi-graded relevance labels based on the self-rated confidence $\text{rating}(q,p)$ that question $q$ is answerable with passage $p$.

For every passage, we identify the highest self-rating that a passage obtained on any exam question in the question bank $Q$. 
\begin{equation}
    \text{relevance-label}(p)=\max_{q\in Q} \text{rating}(q,p)
\end{equation}

This relevance label is then incorporated into the qrels file as a multi-relevance scale. Setting a threshold for multi-graded relevance\footnote{set with \treceval{} option \texttt{-\phantom{}-level\_for\_rel}} considers only those passages as relevant that meet or exceed the specified self-rating level.

% \paragraph{\questionanswering{} with answer verification.}
% \ld{are we still doing this? To avoid lucky guesses, we also experiment with the $k$'t highest rating instead of the maximum.}

% Likewise, when using EXAM with \questionanswering{} and answer verification, the multi-grades relevance label can be derived from the number of correctly answered questions per paragraph.
% \begin{equation}
%     \text{relevance-label}(p)=\left\vert \left\{q \vert \text{correct}(q,p), \forall q\in Q \right\} \right\vert
% \end{equation}

% By adjusting the threshold for multi-graded relevance, we adjust the number of questions that must be answered (at minimum) by a paragraph in order to be counted as relevant.

\bigskip 

While \examqrels{} does not discourage redundant information in the ranking, unlike \examcover{}, it still demonstrates strong rank correlation with official leaderboards in our experiments. This is particularly evident when the official evaluation metrics themselves do not penalize redundant information in the ranking, as observed in the cases of \car{} and \dl{}.

\section{Experimental Evaluation (of the Evaluation Metric)}
\label{sec:evaluation}

\subsection{Experimental Setup}

We study our approach in three experiments.

\begin{description}
\item[\car{} / \tqa{}:] Replicating the experimental setup of Sander et al by using queries and systems from \car{} \citep{dietz2019trec} which we evaluate with exam questions from the \tqa{} dataset \citep{Kembhavi2017TQA}.
\item[\car{} / \genQ{}:] Switching to our prompt-based generated question banks, we demonstrate that similar results are obtained when systems from \car{} are evaluated with a new set of generated questions.
\item[\dl{} 2019 \& 2020 / \genQ{}:] To demonstrate generalization to other datasets we use queries and systems from two years of \dl{} \citep{dl19,dl20} using a generated question bank for each test collection.
\end{description}

\noindent For all four experiments we use rankings of systems submitted to the respective TREC track (test data set) along with manually produced official TREC judgments.

The goals for the evaluation metric are:
\begin{description}
    \item[Leaderboard rank-correlation:] The leaderboard of systems under the \examcover{} and \examqrels{} metric should be similar to the official leaderboard. This is evaluated with two rank correlation measures: Spearman's rank correlation coefficient, measures differences in the leaderboard rank,  and Kendall's $\tau$ rank correlation which penalizes swaps of two systems on the leaderboard.
    \item[Inter-annotator agreement:] High passage-level agreement between official judgments and our predicted relevance labels. We provide count statistics and Cohen's $\kappa$ inter-annotator agreement which corrects for leniency vs strictness.
\end{description}

For question generation we use \texttt{gpt-3.5-turbo-instruct}; for question verification and self-rating we use \texttt{google/flan-t5-large} with the \texttt{text2text-generation} pipeline from Hugging Face.\footnote{\url{https://huggingface.co/google/flan-t5-large}} We also explored fine-tuning the FLAN-T5-large on the \squad{} dataset, but results were slightly worse.

We obtain ten questions for each of the 721 query-subtopics in \car{}, and each query in \dl{}.
We graded all passages in official judgments and the top 20 of all run submissions with EXAM. Across 131 queries in \car{} we grade, 85,329 passages. Across 43 queries in \dlfirst{}, we grade 9,260 passages. Across 54 queries in \dlsecond{}, we grade 11,386 passages. 
 
In \examqrels{} we use \treceval{}'s  \texttt{Precision@20} metric for simplicity. For self-rated approaches, we define a minimum rating of 4 as relevant (``strict'' binary relevance scale) unless otherwise specified.

\textbf{Baselines.}
We demonstrate that our approach is outperforming the original EXAM Answerability Metric \citep{sander2021exam}. Since Sander's work demonstrated that ROUGE metrics are uncorrelated with leaderboard rankings, we omit the comparison here.

Although prompt-based relevance judgments differ significantly in spirit from out approach, we provide several baselines as upper-bound reference  for the \dl{} experiments: \texttt{Sun} \citep{sun2023chatgpt}, \texttt{Fag}  \citep{faggioli2023perspectives}, \texttt{Thom}  (abridged prompt, grade 1 or better counted as relevant) \cite{thomas2023large}. Each prompt is used to directly grade passages for relevance with FLAN-T5-Large. We also explored suggested few-shot prompts but found that all of them lead to a drastically decreased quality---for example, for \texttt{Sun} a Spearman correlation of 0.4 instead of 0.96, and Kendall's $\tau$ of -0.01 instead of 0.24.

\textbf{Significance testing.} We perform a standard-error bar overlap test on all results and only describe significant improvements in the text. The standard error in all experiments ranges between 0.01 and 0.015. For brevity they are omitted here, but are available in the online appendix and illustrated in \ref{fig:car-y3-leaderboard}.

% \ld{move to intro}
% Soft goals:
% \begin{itemize}
%     \item Evaluation method that contributes to a reusable test collection that is repeatable and without biases to systems that did not contribute to the assessment pool. 
%     \item Fully-automatic evaluation method, that permits human verification and intervention.
%     \item Evaluation method that can be extended to improve coverage in areas that were missed.
%     \item Evaluation method that can compare retrieved content, generated content, and retrieval-augmented generated content.
% \end{itemize}

% Hard goals:
% \begin{itemize}
%     \item Evaluation metric that produces leaderboards that are similar to the official leaderboards of the track (but for less manual work)
%     \item (Less critical) Produces relevance labels that have a high inter-annotator agreement with manual judgments.
%     \item (Less critical) Easy integration into existing evaluation software, such as \treceval. 
% \end{itemize}

\subsection{Results on \car{} / \tqa{}}

We first reproduce the experimental setup of the original EXAM method, using the \car{} collection and exam question banks from the \tqa{} dataset. 

As part of the exam grading, the FLAN-T5-large LLM is prompted to answer each exam question with every passage from the system response, which is then verified against the gold standard answer. 

\begin{figure}
    \centering
    \includegraphics[width=1\columnwidth]{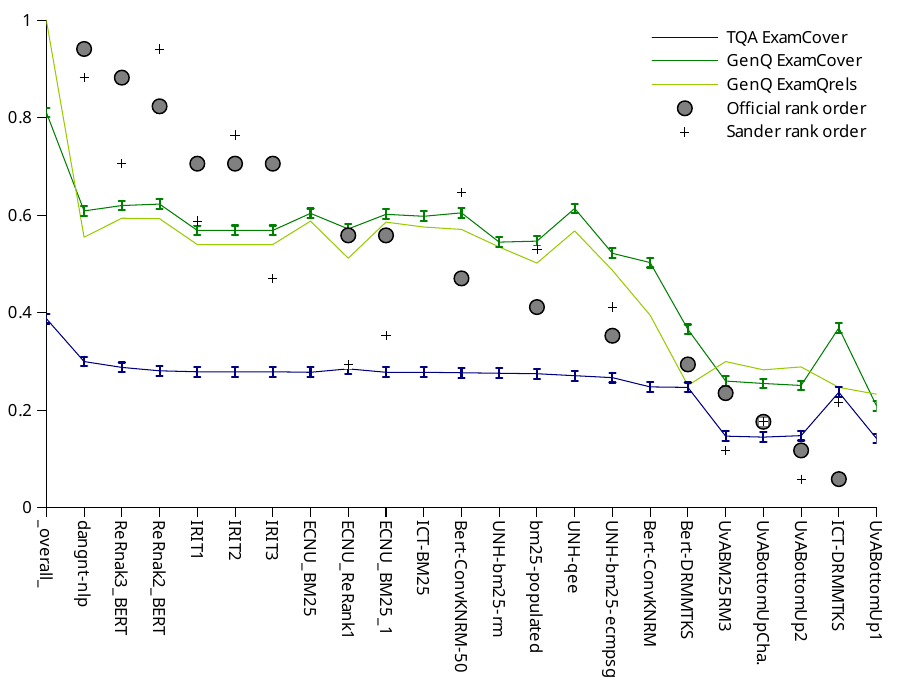}
    \caption{Comparing \car{} leaderboards, our EXAM metrics show strong correlation  with the official TREC leaderboard.    
    Methods are ordered by their official leaderboard rank (filled circles). Methods without official ranks are placed according to their TQA \examcover{} scores. Standard error bars given. For context, Sander's EXAM leaderboard (depicted as crosses) does not show as strong a correlation.}
    \label{fig:car-y3-leaderboard}
\end{figure}

The \examcover{} leaderboard generated using the \tqa{}-based \examcover{} approach is very similar to the official CAR-Y3 leaderboard (Table \ref{tab:car_leaderboard}). With the exception of the \texttt{ECNU\_ReRank1} method, the top 10 of the leaderboard is in exactly the same order. The Spearman rank measure is 0.94, Kendall $\tau$ is 0.84. Our approach outperforms the original EXAM metric significantly, obtaining a correlation comparable to MAP vs NDCG (cf. Table 1 in \citep{sander2021exam}).
This is visually confirmed in Figure \ref{fig:car-y3-leaderboard}.
While several methods obtain similar EXAM scores, these are indeed very similar systems with exactly the same score on the official leaderboard.

Given the distinct nature of our evaluation paradigm, a high inter-annotator agreement on the passage-level is not anticipated. Indeed, as shown in Table \ref{tab:car-tqa-binary-cohen}, the per-passage inter-annotator agreement is very low ($\kappa \approx 0.08$). In contrast to the leaderboard correlation, this demonstrates that a high agreement is not necessary to evaluate system quality correctly.

\begin{table}[]
    \centering
    \caption{\car{} leaderboards, ordered by \examcover{} on TQA questions. In terms of both Spearman's and Kendall's $\tau$ rank correlation coefficients, all our EXAM approaches obtain a better correlation with the official leaderboard than the original EXAM implementation. Note that some submitted systems were not included on the official leaderboard as they had not been assessed.}
    \label{tab:car_leaderboard}
% Preview source code for paragraph 0
\begin{small}

% Preview source code for paragraph 0

\begin{tabular}{lccccc}
\toprule
CAR-Y3 & \makecell{TQA \\ EXAM \\ Cover} & \makecell{GenQ \\  EXAM \\ Cover} & \makecell{GenQ \\ EXAM \\ Qrels} & \makecell{official \\ rank \\ } & \makecell{TQA \\ \citep{sander2021exam} \\ rank}  \tabularnewline
%method && & & & \tabularnewline
% method & \begin{sideways}ExamCover\end{sideways} & \begin{sideways}ExamCover\end{sideways} & \begin{sideways}ExamQrels\end{sideways} & \begin{sideways}rank\end{sideways} & \begin{sideways}rank\end{sideways}\tabularnewline
\midrule 
\_overall\_ & 0.388 & 0.811 & 1 &  & \tabularnewline
dangnt-nlp & 0.300 & 0.609 & 0.555 & 1 & 2\tabularnewline
ReRnak3\_BERT & 0.288 & 0.620 & 0.594 & 2 & 5\tabularnewline
ECNU\_ReRank1 & 0.285 & 0.572 & 0.512 & 8 & 12\tabularnewline
ReRnak2\_BERT & 0.281 & 0.623 & 0.593 & 3 & 1\tabularnewline
IRIT1 & 0.279 & 0.569 & 0.540 & 5 & 7\tabularnewline
IRIT2 & 0.279 & 0.569 & 0.540 & 5 & 4\tabularnewline
IRIT3 & 0.279 & 0.569 & 0.540 & 5 & 9\tabularnewline
ECNU\_BM25 & 0.278 & 0.604 & 0.588 &  & \tabularnewline
ECNU\_BM25\_1 & 0.278 & 0.602 & 0.586 & 8 & 11\tabularnewline
ICT-BM25 & 0.278 & 0.598 & 0.576 &  & \tabularnewline
Bert-ConvKNRM-50 & 0.277 & 0.605 & 0.571 & 9 & 6\tabularnewline
UNH-bm25-rm & 0.276 & 0.545 & 0.535 &  & \tabularnewline
bm25-populated & 0.275 & 0.547 & 0.502 & 10 & 8\tabularnewline
UNH-qee & 0.271 & 0.614 & 0.568 &  & \tabularnewline
UNH-bm25-ecmpsg & 0.267 & 0.522 & 0.487 & 11 & 10\tabularnewline
Bert-ConvKNRM & 0.248 & 0.503 & 0.395 &  & \tabularnewline
Bert-DRMMTKS & 0.247 & 0.366 & 0.251 & 12 & 3\tabularnewline
ICT-DRMMTKS & 0.237 & 0.369 & 0.247 & 16 & 13\tabularnewline
UvABottomUp2 & 0.148 & 0.251 & 0.289 & 15 & 16\tabularnewline
UvABM25RM3 & 0.147 & 0.260 & 0.300 & 13 & 15\tabularnewline
UvABottomUpCh. & 0.145 & 0.255 & 0.283 & 14 & 14\tabularnewline
UvABottomUp1 & 0.142 & 0.209 & 0.233 &  & \tabularnewline
\midrule 
spearman & 0.937 & 0.869 & 0.865 &  & 0.75\tabularnewline
kendall & 0.841 & 0.687 & 0.738 &  & 0.57\tabularnewline
\bottomrule
\end{tabular}
\end{small}

\end{table}

% \normalsize%
% \section{correlation tables}%
% \label{sec:correlationtables}%

% %
% \section{min answers 1}%
% \label{sec:minanswers1}%

%

\begin{table}[]
\caption{Inter-annotator agreement on \car{} / \tqa{} question banks using \questionanswering{} with answer verification. Boxed cells indicate mappings from label to judgment. Largest count per column is depicted in bold.
%min answer = 1
\label{tab:car-tqa-binary-cohen}}
\begin{tabu}{@{}llcclr@{}}%
\toprule%
\multirow{4}{*}{\begin{sideways}BINARY\end{sideways}}&\textbf{Label}&\multicolumn{2}{c}{\textbf{Judgments}}&\textbf{Total}&\textbf{Cohen's }$\boldsymbol{\kappa}$\\%
\cmidrule(l@{\tabcolsep}){3-4}%
&&1+2+3&0&&\\%
\cmidrule(l@{\tabcolsep}){1-6}%
&1&\fbox{{553}}&452&1005&0.078\\%
&0&\textbf{2237}&\fbox{\textbf{3110}}&5347&0.079\\\bottomrule%
\end{tabu}%
\end{table}

\begin{table}[]
\caption{Inter-annotator agreement on \car{} / \genQ{} question banks using self-ratings. Top: Graded ratings-vs-judgments. Rating 4 is most frequently associated with a positive grade. Middle: Lenient relevance for ratings $\geq 1$. Bottom: Strict relevance when ratings  $\geq 4$. %Boxed cells indicate mappings from label to judgment. Largest count per column is depicted in bold.
%min answer = 2
\label{tab:car-genq-cohen}}
\begin{tabu}{@{}llccccr}%
\toprule%
\multirow{8}{*}{\begin{sideways}GRADED\end{sideways}}&\textbf{Label}&\multicolumn{4}{c}{\textbf{Judgments}}&\textbf{Total}\\%
\cmidrule(l@{\tabcolsep}){3-6}%
&&3&2&1&0&\\%
\cmidrule(l@{\tabcolsep}){1-7}%
&5&11&\fbox{{52}}&13&20&96\\%
&4&\textbf{430}&\fbox{\textbf{1121}}&\textbf{472}&{1425}&3448\\%
&3&5&33&\fbox{19}&{98}&155\\%
&2&17&76&\fbox{27}&{217}&337\\%
&1&5&57&\fbox{23}&{160}&245\\%
&0&52&238&139&\fbox{\textbf{1642}}&2071\\\bottomrule%
\end{tabu}%

%with kappa!
% \begin{tabu}{@{}llcccclr@{}}%
% \toprule%
% \multirow{8}{*}{\begin{sideways}GRADED\end{sideways}}&\textbf{Label}&\multicolumn{4}{c}{\textbf{Judgments}}&\textbf{Total}&\textbf{Cohen's }$\boldsymbol{\kappa}$\\%
% \cmidrule(l@{\tabcolsep}){3-6}%
% &&3&2&1&0&&\\%
% \cmidrule(l@{\tabcolsep}){1-8}%
% &5&11&\fbox{{52}}&13&20&96&0.034\\%
% &4&\textbf{430}&\fbox{\textbf{1121}}&\textbf{472}&{1425}&3448&0.17\\%
% &3&5&33&\fbox{19}&{98}&155&0.0059\\%
% &2&17&76&\fbox{27}&{217}&337&{-}0.021\\%
% &1&5&57&\fbox{23}&{160}&245&{-}0.0081\\%
% &0&52&238&139&\fbox{\textbf{1642}}&2071&0.29\\\bottomrule%
% %
% \end{tabu}%

\par%
% \begin{tabu}{@{}llccclr@{}}%
% \toprule%
% \multirow{5}{*}{\begin{sideways}MERGE\end{sideways}}&\textbf{Label}&\multicolumn{3}{c}{\textbf{Judgments}}&\textbf{Total}&\textbf{Cohen's }$\boldsymbol{\kappa}$\\%
% \cmidrule(l@{\tabcolsep}){3-5}%
% &&2+3&1&0&&\\%
% \cmidrule(l@{\tabcolsep}){1-7}%
% &4+5&\fbox{\textbf{1614}}&485&1445&3544&0.28\\%
% &1+2+3&193&\fbox{69}&\textbf{475}&737&{-}0.017\\%
% &0&290&139&\fbox{\textbf{1642}}&2071&0.29\\\bottomrule%
% %
% \end{tabu}%
% \par%

\begin{tabu}{@{}llcclr@{}}%
\toprule%
\multirow{4}{*}{\begin{sideways}LENIENT\end{sideways}}&\textbf{Label}&\multicolumn{2}{c}{\textbf{Judgments}}&\textbf{Total}&\textbf{Cohen's }$\boldsymbol{\kappa}$\\%
\cmidrule(l@{\tabcolsep}){3-4}%
&&1+2+3&0&&\\%
\cmidrule(l@{\tabcolsep}){1-6}%
&1+2+3+4+5&\fbox{\textbf{2361}}&1920&4281&0.3\\%
&0&429&\fbox{\textbf{1642}}&2071&0.29\\\bottomrule%
\end{tabu}%
\par%
\begin{tabu}{@{}llcclr@{}}%
\toprule%
\multirow{4}{*}{\begin{sideways}STRICT\end{sideways}}&\textbf{Label}&\multicolumn{2}{c}{\textbf{Judgments}}&\textbf{Total}&\textbf{Cohen's }$\boldsymbol{\kappa}$\\%
\cmidrule(l@{\tabcolsep}){3-4}%
&&1+2+3&0&&\\%
\cmidrule(l@{\tabcolsep}){1-6}%
&4+5&\fbox{\textbf{2099}}&1445&3544&0.35\\%
&0+1+2+3&691&\fbox{\textbf{2117}}&2808&0.34\\\bottomrule%
\end{tabu}%
\end{table}

\subsection{Results on \car{} with \genQ{}}

When switching to generated question banks, which do not have a correct answer key, we obtain comparable (albeit slightly lower) results. 

The leaderboard correlation (Table \ref{tab:car_leaderboard}, columns marked with \genQ{}) still obtain a very good correlation of Spearman $\approx 0.86$. According to Table \ref{fig:car-y3-leaderboard} (green lines), both the \examcover{} and \examqrels{} methods follow the trend of the official leaderboard. Interestingly, the system ICT-DRMMTKS denotes a positive outlier under both exam question banks.

\subsection{Results on \dl{} with \genQ{}}

Across both years of \dl{} we find that our FLAN-T5-EXAM approach supports the conclusions discussed above. 

We find that the \examqrels{} method correlates exceptionally well with the official leaderboard, obtaining spearman rank correlations of 0.93 in \dlfirst{} and 0.96 on \dlsecond{}, and kendall $\tau$ correlation of 0.8 and 0.84 respectively.  This is slightly better than the relevance-labeling prompt \texttt{Thom} \citep{thomas2023large} and slightlyworse than the prompt \texttt{Sun} \citep{sun2023chatgpt}.

The \examcover{} method correlates slightly less. This is likely because the \examcover{} focuses on coverage more than precision.

As mentioned before, a high inter-annotator agreement is not necessary for good leaderboard correlation. Nevertheless, our EXAM approach obtains an inter-annotator agreement score (Cohen's $\kappa$) that is in the range of relevance-prompting methods.

\begin{table}[]
\caption{Inter-annotator agreement on \dlfirst{} / \genQ{} question banks using self-ratings.  Top: Lenient relevance. Bottom: Strict relevance.
%min answer = 2
\label{tab:dl19-genq-cohen}}
\begin{tabu}{@{}llcclr@{}}%
\toprule%
\multirow{4}{*}{\begin{sideways}LENIENT\end{sideways}}&\textbf{Label}&\multicolumn{2}{c}{\textbf{Judgments}}&\textbf{Total}&\textbf{Cohen's }$\boldsymbol{\kappa}$\\%
\cmidrule(l@{\tabcolsep}){3-4}%
&&2+3&0+1&&\\%
\cmidrule(l@{\tabcolsep}){1-6}%
&1+2+3+4+5&\fbox{\textbf{1627}}&{1809}&3436&0.34\\%
&0&874&\fbox{\textbf{4950}}&5824&0.34\\\bottomrule%
\end{tabu}%
\par%
\begin{tabu}{@{}llcclr@{}}%
\toprule%
\multirow{4}{*}{\begin{sideways}STRICT\end{sideways}}&\textbf{Label}&\multicolumn{2}{c}{\textbf{Judgments}}&\textbf{Total}&\textbf{Cohen's }$\boldsymbol{\kappa}$\\%
\cmidrule(l@{\tabcolsep}){3-4}%
&&2+3&0+1&&\\%
\cmidrule(l@{\tabcolsep}){1-6}%
&4+5&\fbox{\textbf{1439}}&1356&2795&0.36\\%
&0+1+2+3&1062&\fbox{\textbf{5403}}&6465&0.36\\\bottomrule%
\end{tabu}%

Comparison $\kappa$: \texttt{Sun} 0.39 / \texttt{Fag} 0.44 / \texttt{Thom} 0.08
\end{table}

% \subsection{Results on TREC DL 2020 with \genQ{}}
% \section{Min Answers= 2}%
% \label{sec:MinAnswers=2}%
 
% { \color{lightgray}
% %
% \begin{tabu}{@{}llcccclr@{}}%
% \toprule%
% \multirow{8}{*}{\begin{sideways}GRADED\end{sideways}}&\textbf{Label}&\multicolumn{4}{c}{\textbf{Judgments}}&\textbf{Total}&\textbf{Cohen's }$\boldsymbol{\kappa}$\\%
% \cmidrule(l@{\tabcolsep}){3-6}%
% &&3&2&1&0&&\\%
% \cmidrule(l@{\tabcolsep}){1-8}%
% &5&\fbox{37}&70&54&\textbf{162}&323&0.04\\%
% &4&\fbox{368}&554&792&\textbf{1429}&3143&0.11\\%
% &3&37&\fbox{36}&98&\textbf{314}&485&{-}0.011\\%
% &2&21&\fbox{55}&106&\textbf{290}&472&0.018\\%
% &1&5&\fbox{20}&24&\textbf{44}&93&0.021\\%
% &0&178&285&866&\fbox{\textbf{5541}}&6870&0.32\\\bottomrule%
% %
% \end{tabu}%
% \par%
% \begin{tabu}{@{}llccclr@{}}%
% \toprule%
% \multirow{5}{*}{\begin{sideways}MERGE\end{sideways}}&\textbf{Label}&\multicolumn{3}{c}{\textbf{Judgments}}&\textbf{Total}&\textbf{Cohen's }$\boldsymbol{\kappa}$\\%
% \cmidrule(l@{\tabcolsep}){3-5}%
% &&3&2&0+1&&\\%
% \cmidrule(l@{\tabcolsep}){1-7}%
% &4+5&\fbox{405}&624&\textbf{2437}&3466&0.11\\%
% &1+2+3&63&\fbox{111}&\textbf{876}&1050&0.018\\%
% &0&178&285&\fbox{\textbf{6407}}&6870&0.22\\\bottomrule%
% %
% \end{tabu}%
% }

% \par%

\begin{table}[]
\caption{Inter-annotator agreement on \dlsecond{} / \genQ{} question banks using self-ratings.  Top: Lenient relevance. Bottom: Strict relevance.
%min answer = 2
\label{tab:dl20-genq-cohen}}

\begin{tabu}{@{}llcclr@{}}%
\toprule%
\multirow{4}{*}{\begin{sideways}LENIENT\end{sideways}}&\textbf{Label}&\multicolumn{2}{c}{\textbf{Judgments}}&\textbf{Total}&\textbf{Cohen's }$\boldsymbol{\kappa}$\\%
\cmidrule(l@{\tabcolsep}){3-4}%
&&2+3&0+1&&\\%
\cmidrule(l@{\tabcolsep}){1-6}%
&1+2+3+4+5&\fbox{\textbf{1203}}&{3313}&4516&0.22\\%
&0&463&\fbox{\textbf{6407}}&6870&0.22\\\bottomrule%
\end{tabu}%
\par%
\begin{tabu}{@{}llcclr@{}}%
\toprule%
\multirow{4}{*}{\begin{sideways}STRICT\end{sideways}}&\textbf{Label}&\multicolumn{2}{c}{\textbf{Judgments}}&\textbf{Total}&\textbf{Cohen's }$\boldsymbol{\kappa}$\\%
\cmidrule(l@{\tabcolsep}){3-4}%
&&2+3&0+1&&\\%
\cmidrule(l@{\tabcolsep}){1-6}%
&4+5&\fbox{\textbf{1029}}&{2437}&3466&0.25\\%
&0+1+2+3&637&\fbox{\textbf{7283}}&7920&0.25\\\bottomrule%
\end{tabu}%

Comparison $\kappa$: \texttt{Sun} 0.24 / \texttt{Fag} 0.29 / \texttt{Thom} 0.08
\end{table}
% \par%
% Leaderboard correlation on TREC DL 2020

% \subsection{FLAN-T5 family members \ld{drop?}}
% \begin{itemize}
% \item flan-t5-small:   spearman 0.32, kendall 0.29
% \item flan-t5-large: spearman 0.87  kendall 0.69

% \end{itemize}

% min_rating=4 nExam:CorrelationStats(spearman_correlation=0.31532207009587254, kendall_correlation=0.29184818148727604)
% min_rating=4  exam:CorrelationStats(spearman_correlation=0.31532207009587254, kendall_correlation=0.29184818148727604)
 
 % \eject 
 
\subsection{A Worked Example} \label{sec:worked-example}

To illustrate how our FLAN-T5-EXAM method functions, we present a detailed example from the \car{} dataset for query\_id \texttt{tqa2:L\_0384}. The passage presented below was retrieved at rank 1 by the \texttt{dangnt-nlp} method and was assessed by TREC judges as 'MUST be mentioned'.

\begin{quote}

\textbf{Query title}: The Integumentary System \\ \textbf{Query subtopic}: Structure of the Skin\\
\textbf{Passage}:\\
\textbf{ID:} \texttt{b95bf325b7fdacac183b1daf7c118be407f52a3a}

The skin is the largest organ in the human body. Skin is made up of three layers, the epidermis, dermis and the fat layer, also called the hypodermis.  \hl{The epidermis is the outer layer of skin that keeps} vital fluids in and \hl{harmful bacteria out of the body}.  \hl{The dermis is the inner layer of skin} that contains \hl{blood vessels}, \hl{nerves}, hair follicles, oil, and \hl{sweat glands}. Severe damage to large areas of skin exposes the human organism to dehydration and infections that can result in death.\\
\textbf{TREC judgment}: 3 (MUST be mentioned)\\
\end{quote}

% \begin{itemize}
%     \item \textbf{\genQ{}:}
%     \begin{itemize}
%         \item[q1?] How does the epidermis, dermis, and hypodermis work together to provide protection, sensation, and regulation for the body?\newline
%         [EXAM Score] "4"
        
%         % \item[q2?] Can the integumentary system be compromised by diseases and conditions, and if so, how does this impact the health of the skin?\newline
%         % [EXAM Score] "4"
        
%         % \item[q3?] How does the skin act as a barrier against pathogens and other foreign substances?\newline
%         % [EXAM Score] "4"
%     \end{itemize}
    
%     \item \textbf{TQA Questions:}
%     \begin{itemize}
    
%         \item[]NDQ\_007535 Outer layer of the skin \newline
%         [EXAM Score] "4" \newline
%         [Answer] epidermis
%     \end{itemize}
% \end{itemize}

The \tqa{} question \texttt{NDQ\_007535} ``Outer layer of the skin?'' was correctly answered as ``epidermis''. Under the self-rating prompt, FLAN-T5 responds that question can be answered in a mostly relevant way but may have minor gaps (rating 4). 

A generated exam questions, ``How does the epidermis, dermis, and hypodermis work together to provide protection, sensation, and regulation for the body?'', also lead to a rating of 4.
While we cannot automatically confirm the correctness without an answer key, we manually highlight the relevant text spans in the passage.
To incorporate a human-in-the-loop, one would ask FLAN-T5 to produce an answer for the generated question, to be highlighted in a verification interface.

\begin{table}[b]
\caption{Leaderboard for \dlfirst{}, using ratings $\geq 3$ as relevant. 
%Some methods omitted to not violate page margins.
\label{tab:dl2019-leaderboard}}
\begin{tabular}{lllc}
\toprule
DL2019               & \makecell{EXAM\\Cover}  & \makecell{EXAM\\Qrels} & \makecell{Official \\ Rank} \\
% method               &   &   &  \\

\midrule
\_overall\_       & 0.900   & 1      &    \\
idst\_bert\_p2    & 0.707 & 0.759 & 2    \\
idst\_bert\_p3    & 0.707 & 0.756 & 3    \\
p\_exp\_rm3\_bert & 0.700   & 0.755 & 4    \\
idst\_bert\_p1    & 0.702 & 0.752 & 1    \\
idst\_bert\_pr2   & 0.695 & 0.751 & 6    \\
p\_bert           & 0.695 & 0.745 & 5    \\
idst\_bert\_pr1   & 0.688 & 0.745 & 7    \\
p\_exp\_bert      & 0.700   & 0.744 & 8    \\
TUW19-p1-f        & 0.716 & 0.741 & 14   \\
ms\_duet\_passage & 0.684 & 0.740 & 22   \\
runid3            & 0.695 & 0.726 & 12   \\
runid4            & 0.691 & 0.721 & 11   \\
TUW19-p2-f        & 0.714 & 0.720 & 17   \\
srchvrs\_ps\_run2 & 0.712 & 0.713 & 19   \\
TUW19-p3-f        & 0.726 & 0.712 & 13   \\
TUW19-p3-re       & 0.688 & 0.705 & 15   \\
TUW19-p1-re       & 0.686 & 0.704 & 16   \\
TUW19-p2-re       & 0.695 & 0.702 & 20   \\
ICT-CKNRM\_B50    & 0.693 & 0.688 & 23   \\
runid2            & 0.674 & 0.685 & 29   \\
bm25base\_p       & 0.677 & 0.679 & 33   \\
srchvrs\_ps\_run3 & 0.716 & 0.670 & 24   \\
bm25tuned\_prf\_p & 0.665 & 0.659 & 25   \\
bm25tuned\_ax\_p  & 0.665 & 0.651 & 27   \\
srchvrs\_ps\_run1 & 0.681 & 0.644 & 34   \\
ICT-CKNRM\_B      & 0.686 & 0.635 & 21   \\
ICT-BERT2         & 0.686 & 0.635 & 18   \\
test1             & 0.684 & 0.617 & 9    \\
bm25base\_ax\_p   & 0.667 & 0.617 & 26   \\
runid5            & 0.681 & 0.586  & 30   \\
TUA1-1            & 0.684 & 0.585 & 10   \\
bm25tuned\_p      & 0.663 & 0.584 & 35   \\
bm25base\_prf\_p  & 0.660  & 0.145 & 28  \\
% bm25tuned\_rm3\_p    & 0.658 &                 & 31                            \\
% bm25base\_rm3\_p     & 0.658 &                 & 32                            \\
% UNH\_bm25            & 0.647 &                 & 36                            \\
% UNH\_exDL\_bm25      & 0.188 &                 &                               \\
\midrule
                     % &       &                 &                               \\
% min\_rating          & 3     & 3               &     Sun / FagB / HELM                          \\
  &     &          &       \texttt{Sun} / \texttt{Fag} / \texttt{Thom}                          \\
Spearman correlation & 0.72  & 0.93            &    0.96 / 0.96 / 0.86                           \\
Kendall correlation  & 0.54  & 0.80             &    0.85 / 0.85 / 0.67         \\
\bottomrule

\end{tabular}
\end{table}
\begin{table}[]
\caption{Leaderboard for \dlsecond{}, using ratings $\geq 4$ as relevant. Some methods omitted to not violate page margins.\label{tab:dl2020-leaderboard}}
\begin{small}
\begin{tabular}{lllc}
\toprule
DL2020               & \makecell{EXAM\\Cover}  & \makecell{EXAM\\Qrels} & \makecell{Official \\ Rank} \\
% DL2020                & EXAM Cover  & EXAM Qrels & Official Rank \\
% method                &       & Evaluation & leaderboard \\
\midrule
\_overall\_          & 0.911 & 1.000 &    \\
NLE\_pr3             & 0.748 & 0.786 & 12 \\
pash\_f2             & 0.744 & 0.786 & 5  \\
pash\_f3             & 0.741 & 0.786 & 3  \\
pash\_f1             & 0.741 & 0.785 & 4  \\
NLE\_pr2             & 0.737 & 0.783 & 16 \\
CoRT-electra         & 0.731 & 0.781 & 9  \\
pash\_r3             & 0.739 & 0.778 & 1  \\
NLE\_pr1             & 0.743 & 0.777 & 17 \\
p\_d2q\_bm25\_duo    & 0.756 & 0.776 & 6  \\
pinganNLP2           & 0.741 & 0.776 & 13 \\
pinganNLP1           & 0.741 & 0.776 & 15 \\
pash\_r2             & 0.739 & 0.776 & 2  \\
RMIT-Bart            & 0.737 & 0.776 & 10 \\
pash\_r1             & 0.735 & 0.775 & 11 \\
p\_d2q\_rm3\_duo     & 0.752 & 0.773 & 7  \\
rr-pass-roberta      & 0.741 & 0.769 & 22 \\
pinganNLP3           & 0.741 & 0.769 & 14 \\
bigIR-T5-BERT-F      & 0.720 & 0.766 & 26 \\
bcai\_bertl\_pass    & 0.748 & 0.762 & 23 \\
nlm-ens-bst-2        & 0.743 & 0.759 & 28 \\
fr\_pass\_roberta    & 0.733 & 0.758 & 20 \\
bigIR-BERT-R         & 0.728 & 0.758 & 19 \\
bigIR-T5-R           & 0.750 & 0.755 & 24 \\
1                    & 0.728 & 0.755 & 18 \\
p\_bm25rm3\_duo      & 0.739 & 0.754 & 8  \\
nlm-ens-bst-3        & 0.735 & 0.753 & 29 \\
nlm-bert-rr          & 0.722 & 0.750 & 30 \\
bigIR-T5xp-T5-F      & 0.735 & 0.745 & 27 \\
2                    & 0.730 & 0.744 & 25 \\
bigIR-DCT-T5-F       & 0.733 & 0.741 & 21 \\
relemb\_mlm\_0\_2    & 0.724 & 0.739 & 31 \\
nlm-prfun-bert       & 0.730 & 0.728 & 32 \\
TUW-TK-2Layer        & 0.739 & 0.722 & 34 \\
p\_d2q\_bm25         & 0.713 & 0.699 & 35 \\
p\_d2q\_bm25rm3      & 0.693 & 0.693 & 36 \\
bert\_6              & 0.726 & 0.691 & 37 \\
TUW-TK-Sparse        & 0.700 & 0.682 & 33 \\
CoRT-bm25            & 0.694 & 0.664 & 38 \\
indri-fdm            & 0.709 & 0.652 & 43 \\
bl\_bcai\_mdl1\_vt   & 0.707 & 0.650 & 40 \\
terrier-DPH          & 0.689 & 0.650 & 52 \\
indri-lmds           & 0.702 & 0.649 & 47 \\
terrier-InL2         & 0.693 & 0.649 & 44 \\
bcai\_class\_pass    & 0.694 & 0.647 & 41 \\
indri-sdm            & 0.698 & 0.642 & 48 \\
terrier-BM25         & 0.683 & 0.641 & 45 \\
DLH\_d\_5\_t\_25     & 0.650 & 0.641 & 46 \\
% bl\_bcai\_mdl1\_vs   & 0.681 & 0.636 & 42 \\
% CoRT-standalone      & 0.676 & 0.630 & 39 \\
% p\_bm25rm3           & 0.637 & 0.625 & 49 \\
% TF\_IDF\_d\_2\_t\_50 & 0.663 & 0.619 & 53 \\
% p\_bm25              & 0.650 & 0.598 & 50 \\
% bm25\_bert\_token    & 0.667 & 0.585 & 51 \\
% small\_1k            & 0.635 & 0.490 & 54 \\
% med\_1k              & 0.646 & 0.474 & 55 \\
% DoRA\_Large\_1k      & 0.633 & 0.468 & 56 \\
 $\dots$ & $\dots$ & $\dots$& $\dots$ \\
DoRA\_Small          & 0.285 & 0.182 & 57 \\
DoRA\_Large          & 0.278 & 0.171 & 59 \\
DoRA\_Med            & 0.270 & 0.171 & 58 \\
\midrule
  &     &          &       \texttt{Sun} / \texttt{Fag} / \texttt{Thom}                          \\
Spearman Correlation & 0.88  & 0.96      &        0.96 / 0.95 / 0.95                      \\
Kendall Correlation  & 0.70   & 0.84      &        0.88 / 0.85 / 0.84 \\     
\bottomrule
\end{tabular}
\end{small}
\end{table}

\section{conclusion}
\label{sec:conclusion}

With FLAN-T5-EXAM we are proposing an alternative evaluation approach that does not make any use of manual passage-level relevance judgments---and does not attempt to mimic relevance judgment process.
Instead, an exam question bank is created as part of topic development, envisioning that each question addresses one important piece of information content. As a result, whenever such questions are answerable with text from an \irsystem{}, we conclude that the system provides relevant information. 

Using three TREC data sets, we demonstrate that (1) our proposed approach can reproduce official TREC leaderboards nearly perfectly; and (2) it is a strong contender in comparison to other recent LLM-based relevance label predictors \citep{sun2023chatgpt,faggioli2023perspectives,thomas2023large}. In contrast, FLAN-T5-EXAM offers a clear path towards integrating a human-in-the-loop, by supporting the refinement of the exam question banks, which is how relevant information content is defined.

We believe that further research will improve the question bank generation, as well as study the positive effects of this approach on the quality, cost, and satisfaction of human judges.

We hope that by integrating EXAM evaluation metric into trec\_eval, we offer a system that can be easily adopted by future IR evaluation tracks, offering organizers an avenue to reduce assessment costs, obtain reusable test collections for generative information systems.

\ifnum\value{UseArxiv}=1
\begin{acks}
This material is based upon work supported by the National Science Foundation under
    Grant No. 1846017. Any opinions, findings, and conclusions or recommendations expressed in this material
    are those of the author(s) and do not necessarily reflect the views of the National Science
    Foundation.

\end{acks}
\else
\fi

\bibliographystyle{ACM-Reference-Format}
\bibliography{bibliography}


\begin{thebibliography}{28}

%%% the Web address.

\ifx \showCODEN    \undefined \def \showCODEN     #1{\unskip}     \fi
\ifx \showDOI      \undefined \def \showDOI       #1{#1}\fi
\ifx \showISBNx    \undefined \def \showISBNx     #1{\unskip}     \fi
\ifx \showISBNxiii \undefined \def \showISBNxiii  #1{\unskip}     \fi
\ifx \showISSN     \undefined \def \showISSN      #1{\unskip}     \fi
\ifx \showLCCN     \undefined \def \showLCCN      #1{\unskip}     \fi
\ifx \shownote     \undefined \def \shownote      #1{#1}          \fi
\ifx \showarticletitle \undefined \def \showarticletitle #1{#1}   \fi
\ifx \showURL      \undefined \def \showURL       {\relax}        \fi
% invisible to TeX
\providecommand\bibfield[2]{#2}
\providecommand\bibinfo[2]{#2}
\providecommand\natexlab[1]{#1}
\providecommand\showeprint[2][]{arXiv:#2}

\bibitem[Arabzadeh et~al\mbox{.}(2024)]%
        {arabzadeh2024adapting}
\bibfield{author}{\bibinfo{person}{Negar Arabzadeh}, \bibinfo{person}{Amin
  Bigdeli}, {and} \bibinfo{person}{Charles~LA Clarke}.}
  \bibinfo{year}{2024}\natexlab{}.
\newblock \showarticletitle{Adapting Standard Retrieval Benchmarks to Evaluate
  Generated Answers}.
\newblock \bibinfo{journal}{\emph{arXiv preprint arXiv:2401.04842}}
  (\bibinfo{year}{2024}).
\newblock


\bibitem[Chang et~al\mbox{.}(2023)]%
        {chang2023survey}
\bibfield{author}{\bibinfo{person}{Yupeng Chang}, \bibinfo{person}{Xu Wang},
  \bibinfo{person}{Jindong Wang}, \bibinfo{person}{Yuan Wu},
  \bibinfo{person}{Kaijie Zhu}, \bibinfo{person}{Hao Chen},
  \bibinfo{person}{Linyi Yang}, \bibinfo{person}{Xiaoyuan Yi},
  \bibinfo{person}{Cunxiang Wang}, \bibinfo{person}{Yidong Wang},
  {et~al\mbox{.}}} \bibinfo{year}{2023}\natexlab{}.
\newblock \showarticletitle{A survey on evaluation of large language models}.
\newblock \bibinfo{journal}{\emph{arXiv preprint arXiv:2307.03109}}
  (\bibinfo{year}{2023}).
\newblock


\bibitem[Clarke and Lapata(2010)]%
        {clarke2010discourse}
\bibfield{author}{\bibinfo{person}{James Clarke} {and} \bibinfo{person}{Mirella
  Lapata}.} \bibinfo{year}{2010}\natexlab{}.
\newblock \showarticletitle{Discourse Constraints for Document Compression}.
\newblock \bibinfo{journal}{\emph{Computational Linguistics}}
  \bibinfo{volume}{36}, \bibinfo{number}{3} (\bibinfo{year}{2010}).
\newblock


\bibitem[Craswell et~al\mbox{.}(2021)]%
        {dl20}
\bibfield{author}{\bibinfo{person}{Nick Craswell}, \bibinfo{person}{Bhaskar
  Mitra}, \bibinfo{person}{Emine Yilmaz}, {and} \bibinfo{person}{Daniel
  Campos}.} \bibinfo{year}{2021}\natexlab{}.
\newblock \showarticletitle{Overview of the TREC 2020 deep learning track}.
\newblock \bibinfo{journal}{\emph{arXiv preprint arXiv:2102.07662}}
  (\bibinfo{year}{2021}).
\newblock


\bibitem[Craswell et~al\mbox{.}(2020)]%
        {dl19}
\bibfield{author}{\bibinfo{person}{Nick Craswell}, \bibinfo{person}{Bhaskar
  Mitra}, \bibinfo{person}{Emine Yilmaz}, \bibinfo{person}{Daniel Campos},
  {and} \bibinfo{person}{Ellen~M Voorhees}.} \bibinfo{year}{2020}\natexlab{}.
\newblock \showarticletitle{Overview of the TREC 2019 deep learning track}.
\newblock \bibinfo{journal}{\emph{arXiv preprint arXiv:2003.07820}}
  (\bibinfo{year}{2020}).
\newblock


\bibitem[Deutsch et~al\mbox{.}(2020)]%
        {deutsch2020questionanswering}
\bibfield{author}{\bibinfo{person}{Daniel Deutsch}, \bibinfo{person}{Tania
  Bedrax-Weiss}, {and} \bibinfo{person}{Dan Roth}.}
  \bibinfo{year}{2020}\natexlab{}.
\newblock \showarticletitle{Towards Question-Answering as an Automatic Metric
  for Evaluating the Content Quality of a Summary}.
\newblock \bibinfo{journal}{\emph{arXiv preprint arXiv:2010.00490}}
  (\bibinfo{year}{2020}).
\newblock


\bibitem[Dietz and Foley(2019)]%
        {dietz2019trec}
\bibfield{author}{\bibinfo{person}{Laura Dietz} {and} \bibinfo{person}{John
  Foley}.} \bibinfo{year}{2019}\natexlab{}.
\newblock \showarticletitle{TREC CAR Y3: Complex Answer Retrieval Overview}. In
  \bibinfo{booktitle}{\emph{Proceedings of {Text REtrieval Conference}
  (TREC)}}.
\newblock


\bibitem[Eyal et~al\mbox{.}(2019)]%
        {eyal-etal-2019-question-summary-first}
\bibfield{author}{\bibinfo{person}{Matan Eyal}, \bibinfo{person}{Tal Baumel},
  {and} \bibinfo{person}{Michael Elhadad}.} \bibinfo{year}{2019}\natexlab{}.
\newblock \showarticletitle{Question Answering as an Automatic Evaluation
  Metric for News Article Summarization}. In
  \bibinfo{booktitle}{\emph{Proceedings of the 2019 Conference of the North
  {A}merican Chapter of the Association for Computational Linguistics: Human
  Language Technologies, Volume 1 (Long and Short Papers)}}.
  \bibinfo{publisher}{Association for Computational Linguistics},
  \bibinfo{address}{Minneapolis, Minnesota}, \bibinfo{pages}{3938--3948}.
\newblock
\urldef\tempurl%
\url{https://doi.org/10.18653/v1/N19-1395}
\showDOI{\tempurl}


\bibitem[Faggioli et~al\mbox{.}(2023)]%
        {faggioli2023perspectives}
\bibfield{author}{\bibinfo{person}{Guglielmo Faggioli}, \bibinfo{person}{Laura
  Dietz}, \bibinfo{person}{Charles~LA Clarke}, \bibinfo{person}{Gianluca
  Demartini}, \bibinfo{person}{Matthias Hagen}, \bibinfo{person}{Claudia
  Hauff}, \bibinfo{person}{Noriko Kando}, \bibinfo{person}{Evangelos Kanoulas},
  \bibinfo{person}{Martin Potthast}, \bibinfo{person}{Benno Stein},
  {et~al\mbox{.}}} \bibinfo{year}{2023}\natexlab{}.
\newblock \showarticletitle{Perspectives on large language models for relevance
  judgment}. In \bibinfo{booktitle}{\emph{Proceedings of the 2023 ACM SIGIR
  International Conference on Theory of Information Retrieval}}.
  \bibinfo{pages}{39--50}.
\newblock


\bibitem[Fok and Weld(2023)]%
        {fok2023search}
\bibfield{author}{\bibinfo{person}{Raymond Fok} {and} \bibinfo{person}{Daniel~S
  Weld}.} \bibinfo{year}{2023}\natexlab{}.
\newblock \showarticletitle{In Search of Verifiability: Explanations Rarely
  Enable Complementary Performance in AI-Advised Decision Making}.
\newblock \bibinfo{journal}{\emph{arXiv preprint arXiv:2305.07722}}
  (\bibinfo{year}{2023}).
\newblock


\bibitem[Huang et~al\mbox{.}(2020)]%
        {Huang2020qasummary}
\bibfield{author}{\bibinfo{person}{Luyang Huang}, \bibinfo{person}{Lingfei Wu},
  {and} \bibinfo{person}{Lu Wang}.} \bibinfo{year}{2020}\natexlab{}.
\newblock \showarticletitle{Knowledge Graph-Augmented Abstractive Summarization
  with Semantic-Driven Cloze Reward}.
\newblock \bibinfo{journal}{\emph{Proceedings of the 58th Annual Meeting of the
  Association for Computational Linguistics}} (\bibinfo{year}{2020}).
\newblock
\urldef\tempurl%
\url{https://doi.org/10.18653/v1/2020.acl-main.457}
\showDOI{\tempurl}


\bibitem[Kembhavi et~al\mbox{.}(2017)]%
        {Kembhavi2017TQA}
\bibfield{author}{\bibinfo{person}{Aniruddha Kembhavi},
  \bibinfo{person}{Minjoon Seo}, \bibinfo{person}{Dustin Schwenk},
  \bibinfo{person}{Jonghyun Choi}, \bibinfo{person}{Ali Farhadi}, {and}
  \bibinfo{person}{Hannaneh Hajishirzi}.} \bibinfo{year}{2017}\natexlab{}.
\newblock \showarticletitle{Are You Smarter Than a Sixth Grader? {T}extbook
  Question Answering for Multimodal Machine Comprehension}.
\newblock \bibinfo{journal}{\emph{2017 IEEE Conference on Computer Vision and
  Pattern Recognition (CVPR)}} (\bibinfo{year}{2017}),
  \bibinfo{pages}{5376--5384}.
\newblock


\bibitem[Li et~al\mbox{.}(2023)]%
        {li2023llatrieval}
\bibfield{author}{\bibinfo{person}{Xiaonan Li}, \bibinfo{person}{Changtai Zhu},
  \bibinfo{person}{Linyang Li}, \bibinfo{person}{Zhangyue Yin},
  \bibinfo{person}{Tianxiang Sun}, {and} \bibinfo{person}{Xipeng Qiu}.}
  \bibinfo{year}{2023}\natexlab{}.
\newblock \showarticletitle{LLatrieval: LLM-Verified Retrieval for Verifiable
  Generation}.
\newblock \bibinfo{journal}{\emph{arXiv preprint arXiv:2311.07838}}
  (\bibinfo{year}{2023}).
\newblock


\bibitem[Liang et~al\mbox{.}(2022)]%
        {liang2022holistic}
\bibfield{author}{\bibinfo{person}{Percy Liang}, \bibinfo{person}{Rishi
  Bommasani}, \bibinfo{person}{Tony Lee}, \bibinfo{person}{Dimitris Tsipras},
  \bibinfo{person}{Dilara Soylu}, \bibinfo{person}{Michihiro Yasunaga},
  \bibinfo{person}{Yian Zhang}, \bibinfo{person}{Deepak Narayanan},
  \bibinfo{person}{Yuhuai Wu}, \bibinfo{person}{Ananya Kumar}, {et~al\mbox{.}}}
  \bibinfo{year}{2022}\natexlab{}.
\newblock \showarticletitle{Holistic evaluation of language models}.
\newblock \bibinfo{journal}{\emph{arXiv preprint arXiv:2211.09110}}
  (\bibinfo{year}{2022}).
\newblock


\bibitem[Lin and Demner-Fushman(2006)]%
        {lin2006will}
\bibfield{author}{\bibinfo{person}{Jimmy Lin} {and} \bibinfo{person}{Dina
  Demner-Fushman}.} \bibinfo{year}{2006}\natexlab{}.
\newblock \showarticletitle{Will pyramids built of nuggets topple over?}. In
  \bibinfo{booktitle}{\emph{Proceedings of the Human Language Technology
  Conference of the NAACL, Main Conference}}. \bibinfo{pages}{383--390}.
\newblock


\bibitem[Longpre et~al\mbox{.}(2023)]%
        {longpre2023flan}
\bibfield{author}{\bibinfo{person}{Shayne Longpre}, \bibinfo{person}{Le Hou},
  \bibinfo{person}{Tu Vu}, \bibinfo{person}{Albert Webson},
  \bibinfo{person}{Hyung~Won Chung}, \bibinfo{person}{Yi Tay},
  \bibinfo{person}{Denny Zhou}, \bibinfo{person}{Quoc~V Le},
  \bibinfo{person}{Barret Zoph}, \bibinfo{person}{Jason Wei}, {et~al\mbox{.}}}
  \bibinfo{year}{2023}\natexlab{}.
\newblock \showarticletitle{The flan collection: Designing data and methods for
  effective instruction tuning}.
\newblock \bibinfo{journal}{\emph{arXiv preprint arXiv:2301.13688}}
  (\bibinfo{year}{2023}).
\newblock


\bibitem[MacAvaney and Soldaini(2023)]%
        {macavaney2023one}
\bibfield{author}{\bibinfo{person}{Sean MacAvaney} {and} \bibinfo{person}{Luca
  Soldaini}.} \bibinfo{year}{2023}\natexlab{}.
\newblock \showarticletitle{One-Shot Labeling for Automatic Relevance
  Estimation}.
\newblock \bibinfo{journal}{\emph{arXiv preprint arXiv:2302.11266}}
  (\bibinfo{year}{2023}).
\newblock


\bibitem[McCreadie and Buntain(2023)]%
        {mccreadie2023crisisfacts}
\bibfield{author}{\bibinfo{person}{Richard McCreadie} {and}
  \bibinfo{person}{Cody Buntain}.} \bibinfo{year}{2023}\natexlab{}.
\newblock \bibinfo{booktitle}{\emph{CrisisFACTS: Buidling and Evaluating Crisis
  Timelines}}.
\newblock \bibinfo{type}{{T}echnical {R}eport}. \bibinfo{institution}{Univerity
  of Glasgow}.
\newblock


\bibitem[Mulla and Gharpure(2023)]%
        {mulla2023automatic}
\bibfield{author}{\bibinfo{person}{Nikahat Mulla} {and} \bibinfo{person}{Prachi
  Gharpure}.} \bibinfo{year}{2023}\natexlab{}.
\newblock \showarticletitle{Automatic question generation: a review of
  methodologies, datasets, evaluation metrics, and applications}.
\newblock \bibinfo{journal}{\emph{Progress in Artificial Intelligence}}
  \bibinfo{volume}{12}, \bibinfo{number}{1} (\bibinfo{year}{2023}),
  \bibinfo{pages}{1--32}.
\newblock


\bibitem[Pavlu et~al\mbox{.}(2012)]%
        {pavlu2012ir}
\bibfield{author}{\bibinfo{person}{Virgil Pavlu}, \bibinfo{person}{Shahzad
  Rajput}, \bibinfo{person}{Peter~B Golbus}, {and} \bibinfo{person}{Javed~A
  Aslam}.} \bibinfo{year}{2012}\natexlab{}.
\newblock \showarticletitle{IR system evaluation using nugget-based test
  collections}. In \bibinfo{booktitle}{\emph{Proceedings of the fifth ACM
  international conference on Web search and data mining}}.
  \bibinfo{pages}{393--402}.
\newblock


\bibitem[Sachan et~al\mbox{.}(2022)]%
        {sachan2022improving}
\bibfield{author}{\bibinfo{person}{Devendra~Singh Sachan},
  \bibinfo{person}{Mike Lewis}, \bibinfo{person}{Mandar Joshi},
  \bibinfo{person}{Armen Aghajanyan}, \bibinfo{person}{Wen-tau Yih},
  \bibinfo{person}{Joelle Pineau}, {and} \bibinfo{person}{Luke Zettlemoyer}.}
  \bibinfo{year}{2022}\natexlab{}.
\newblock \showarticletitle{Improving passage retrieval with zero-shot question
  generation}.
\newblock \bibinfo{journal}{\emph{arXiv preprint arXiv:2204.07496}}
  (\bibinfo{year}{2022}).
\newblock


\bibitem[Sander and Dietz(2021)]%
        {sander2021exam}
\bibfield{author}{\bibinfo{person}{David~P Sander} {and} \bibinfo{person}{Laura
  Dietz}.} \bibinfo{year}{2021}\natexlab{}.
\newblock \showarticletitle{EXAM: How to Evaluate Retrieve-and-Generate Systems
  for Users Who Do Not (Yet) Know What They Want.}. In
  \bibinfo{booktitle}{\emph{DESIRES}}. \bibinfo{pages}{136--146}.
\newblock


\bibitem[Sun et~al\mbox{.}(2023)]%
        {sun2023chatgpt}
\bibfield{author}{\bibinfo{person}{Weiwei Sun}, \bibinfo{person}{Lingyong Yan},
  \bibinfo{person}{Xinyu Ma}, \bibinfo{person}{Pengjie Ren},
  \bibinfo{person}{Dawei Yin}, {and} \bibinfo{person}{Zhaochun Ren}.}
  \bibinfo{year}{2023}\natexlab{}.
\newblock \showarticletitle{Is ChatGPT Good at Search? Investigating Large
  Language Models as Re-Ranking Agent}.
\newblock \bibinfo{journal}{\emph{arXiv e-prints}} (\bibinfo{year}{2023}),
  \bibinfo{pages}{arXiv--2304}.
\newblock


\bibitem[Thomas et~al\mbox{.}(2023)]%
        {thomas2023large}
\bibfield{author}{\bibinfo{person}{Paul Thomas}, \bibinfo{person}{Seth
  Spielman}, \bibinfo{person}{Nick Craswell}, {and} \bibinfo{person}{Bhaskar
  Mitra}.} \bibinfo{year}{2023}\natexlab{}.
\newblock \bibinfo{title}{Large language models can accurately predict searcher
  preferences}.
\newblock
\newblock
\showeprint[arxiv]{2309.10621}~[cs.IR]


\bibitem[Wang et~al\mbox{.}(2020)]%
        {wang-etal-2020-factual-qa}
\bibfield{author}{\bibinfo{person}{Alex Wang}, \bibinfo{person}{Kyunghyun Cho},
  {and} \bibinfo{person}{Mike Lewis}.} \bibinfo{year}{2020}\natexlab{}.
\newblock \showarticletitle{Asking and Answering Questions to Evaluate the
  Factual Consistency of Summaries}. In \bibinfo{booktitle}{\emph{Proceedings
  of the 58th Annual Meeting of the Association for Computational
  Linguistics}}. \bibinfo{publisher}{Association for Computational
  Linguistics}, \bibinfo{address}{Online}, \bibinfo{pages}{5008--5020}.
\newblock
\urldef\tempurl%
\url{https://doi.org/10.18653/v1/2020.acl-main.450}
\showDOI{\tempurl}


\bibitem[Wang et~al\mbox{.}(2023)]%
        {wang2023large}
\bibfield{author}{\bibinfo{person}{Peiyi Wang}, \bibinfo{person}{Lei Li},
  \bibinfo{person}{Liang Chen}, \bibinfo{person}{Dawei Zhu},
  \bibinfo{person}{Binghuai Lin}, \bibinfo{person}{Yunbo Cao},
  \bibinfo{person}{Qi Liu}, \bibinfo{person}{Tianyu Liu}, {and}
  \bibinfo{person}{Zhifang Sui}.} \bibinfo{year}{2023}\natexlab{}.
\newblock \showarticletitle{Large language models are not fair evaluators}.
\newblock \bibinfo{journal}{\emph{arXiv preprint arXiv:2305.17926}}
  (\bibinfo{year}{2023}).
\newblock


\bibitem[Weng et~al\mbox{.}(2023)]%
        {weng2023large}
\bibfield{author}{\bibinfo{person}{Yixuan Weng}, \bibinfo{person}{Fei Xia},
  \bibinfo{person}{Bin Li}, \bibinfo{person}{Shizhu He},
  \bibinfo{person}{Shengping Liu}, \bibinfo{person}{Bin Sun},
  \bibinfo{person}{Kang Liu}, \bibinfo{person}{Jun Zhao}, {et~al\mbox{.}}}
  \bibinfo{year}{2023}\natexlab{}.
\newblock \showarticletitle{Large Language Models are Better Reasoners with
  Self-Verification}. In \bibinfo{booktitle}{\emph{The 2023 Conference on
  Empirical Methods in Natural Language Processing}}.
\newblock


\bibitem[Zhang and Gao(2023)]%
        {zhang2023towards}
\bibfield{author}{\bibinfo{person}{Xuan Zhang} {and} \bibinfo{person}{Wei
  Gao}.} \bibinfo{year}{2023}\natexlab{}.
\newblock \showarticletitle{Towards llm-based fact verification on news claims
  with a hierarchical step-by-step prompting method}.
\newblock \bibinfo{journal}{\emph{arXiv preprint arXiv:2310.00305}}
  (\bibinfo{year}{2023}).
\newblock


\end{thebibliography}

\end{document}